
\input epsf
\magnification\magstep1



\hbadness=10000      
\vbadness=10000  

\font\eightit=cmti8  
\font\eightrm=cmr8 \font\eighti=cmmi8                 
\font\eightsy=cmsy8 
           \font\sixrm=cmr6



\def\eightpoint{\normalbaselineskip=10pt 
\def\rm{\eightrm\fam0} \let\it\eightit
\textfont0=\eightrm \scriptfont0=\sixrm 
\textfont1=\eighti \scriptfont1=\seveni
\textfont2=\eightsy \scriptfont2=\sevensy 
\normalbaselines \eightrm
\parindent=1em}



\def\eq#1{{\noexpand\rm(#1)}}          
\newcount\eqcounter                    
\eqcounter=0                           
\def\numeq{\global\advance\eqcounter by 1\eq{\the\eqcounter}}           
\def\relativeq#1{{\advance\eqcounter by #1\eq{\the\eqcounter}}}


\def\namelasteq#1{\global\edef#1{{\eq{\the\eqcounter}}}}  

\def\A{{\rm A}}                        
\def\Amu{{\rm A}_\mu}
\def\Arho{{\rm A}_\rho}
\def\Anu{{\rm A}_\nu}
\def\barc{{\bar{\rm c}}}
\def\bars{{\bar s}}

\def\E{{\rm E}}
\def\levicivita{\epsilon^{\mu\rho\nu}}
\def\measurex{\int\!{d^3 x}}
\def\measureq{\int\!{d^3 q\over (2\pi)^3}}
\def\partirho{\partial_{\rho}}
\def\rb{{\rm b}}
\def\rc{{\rm c}}
\def\rD{{\rm D}}
\def\Reg{{\rm R}(-{\partial^2\over\Lambda^2})}

\def\cite#1{{\rm[#1]}}                 
\def\Trace{{\rm Tr}}



\newif\ifstartsec                   

\outer\def\section#1{\vskip 0pt plus .15\vsize \penalty -250
\vskip 0pt plus -.15\vsize \bigskip \startsectrue
\message{#1}\centerline{\bf#1}\nobreak\noindent}

\def\subsection#1{\ifstartsec\medskip\else\bigskip\fi \startsecfalse
\noindent{\it#1}\penalty100\medskip}

\def\refno#1. #2\par{\smallskip\item{\rm\lbrack#1\rbrack}#2\par}

\hyphenation{geo-me-try}

\bigskip


\def\TMYM{1}
\def\Wittenjones{2}
\def\HALL{3}
\def\Labastida{4}
\def\Vassinv{5}
\def\King{6}
\def\PP{7}
\def\Krajewski{8}
\def\NCCSper{9}
\def\NCCSnoper{10}
\def\integerk{11}
\def\induced{12}
\def\NCHALL{13}
\def\Piguet{14}
\def\cpmaxial{15}
\def\Power{16}
\def\NCWick{17}
\def\LM{18}
\def\GLM{19}
\def\axialbooks{20}
\def\Leibmart{21}
\def\Leibchen{22}
\def\WZW{23}
\def\DasJab{24}


\rightline{FT/UCM--80--2001}

\vskip 1cm

\centerline{\bf Computing noncommutative Chern-Simons theory 
radiative corrections}
\centerline{\bf on the back of an envelope}

\bigskip

\centerline{\rm C. P. Mart{\'\i}n\dag}
\medskip
\centerline{\eightit Departamento de F{\'\i}sica Te\'orica I,						Universidad Complutense, 28040 Madrid, Spain}
\vfootnote\dag{E-mail: {\tt carmelo@elbereth.fis.ucm.es}}

\bigskip
\rightline{\it In memory of George Leibbrandt}
 \vskip 0.5 truecm 

\begingroup\narrower\narrower
\eightpoint
We show that the renormalized $U(N)$ noncommutative Chern-Simons theory can be 
defined in perturbation theory so that there are no loop corrections to 
the 1PI functional of the theory in an arbitrary homogeneous axial (time-like, 
light-like or space-like) gauge. We define the free propagators of the fields 
of the theory by using the Leibbrandt-Mandelstam prescription --which allows 
Wick rotation and is consistent with power-counting-- and regularize 
its Green functions with the help of a family of regulators which explicitly 
preserve the infinitesimal vector Grassmann symmetry of the theory. We also 
show that in perturbation theory the nonvanishing Green functions of the 
elementary fields of the theory  are products of the free propagators.

\par
\endgroup 
\vskip 1cm

\section{1.- Introduction}

In 1982~\cite{\TMYM}, the integral of the Chern-Simons three-form was added 
to the Yang-Mills action  to give the gauge field in three dimensions a mass 
without the need of the Higgs mechanism. Thus Topologically Massive Yang-Mills
theory was born. In 1988, Nonabelian Chern-Simons theory~\cite{\Wittenjones} 
--the theory whose action is the integral over a given (commutative) 
three-dimensional manifold of the Chern-Simons three form for a nonabelian 
gauge group-- was introduced to give a three-dimensional and 
field theoretical definition of the Jones polynomial and other topological 
invariants of three-dimensional manifolds. When the gauge 
field of the Chern-Simons action couples to matter fields we obtain field 
theories having anyonic excitations, these field theories  being relevant
in the long wavelength range description of the fractional quantum Hall effect 
and high temperature conductivity~\cite{\HALL}.

The perturbative construction of Nonabelian Chern-Simons theory has been the
subject of intensive research and has yielded very beautiful and deep results 
--see ref.~\cite{\Labastida} and references therein. In particular, the 
perturbative study of the observables of the theory in the temporal and 
light-cone gauges  has given rise~\cite{\Vassinv}   
respectively to a combinatorial 
representation of the Vassiliev invariants and the representation of these 
invariants  furnished by the Kontsevich integral. The Light-cone gauge 
had been used much earlier~\cite{\King} to quantize Nonabelian Chen-Simons 
theory and obtain the Knizhnik-Zamolodchikov equation as the differential 
equation to be satisfied by the Wilson line vevs, thus  making connection 
with the Wess-Zumino-Witten model. 

The Chern-Simons action over noncommutative manifolds~\cite{\PP} --henceforth 
the noncommutative Chern-Simons action--  was first considered in 
refs.~\cite{\Krajewski}. The properties of Chern-Simons theory on the  
noncommutative plane --henceforth  noncommutative Chern-Simons theory-- has 
been studied perturbatively~\cite{\NCCSper} and 
nonperturbatively~\cite{\NCCSnoper, \integerk} in a number of papers. 
The noncommutative Chern-Simons action as an induced action has been 
considered in ref.~\cite{\induced}. Fractional  quantum Hall fluids can be 
given an effective description by using the noncommutative Chern-Simons 
theory~\cite{\NCHALL}. It would appear that noncommutative Chern-Simons 
theory will play in Physics as  outstanding a role as its ordinary  
sibling does.  
 
The purpose of this paper is to quantize perturbatively the noncommutative 
Chern-Simons theory in a general homogeneous axial-type gauge  
--$n^{\mu}\,\Amu =0$, either with $n^2 > 0$ (space-like),  $n^2 = 0$ (light-cone), or with $n^2 < 0$ (time-like). We shall use the generalized Leibbrandt-Mandelstam prescription to go around the axial infrared singularity of the naive 
propagators and thus treat on equal footing the three types of axial gauges. 
By employing a rather arbitrary regularization method  which explicitly 
preserves the Grassmann vector symmetry of the theory, we shall be able to 
compute, at any order in perturbation theory,  the radiative corrections 
to Green functions of the elementary fields. We shall thus show  that in the
theory so defined these Green functions  lack radiative  corrections and 
that they are, the Green functions, either products of the free propagators or 
vanish.    

The layout of this paper is as follows. In section 2, we establish the 
symmetries of the gauge fixed theory and perform the computation of the 
quantum corrections to the 1PI functional. Section 3 contains the conclusions 
and some comments.

\section{2.- Symmetries and quantum corrections}

To define in perturbation theory the noncommutative $U(N)$ Chern-Simons theory in an homogeneous axial type gauge  one defines, perturbatively, the path 
integral with the help of the following action
$$ 
 S=-{k\over 4\pi}\Trace\measurex\,\levicivita(\Amu\star\partirho\Anu
-{2i \over 3}\Amu\star\Arho\star\Anu)\,+\,2\,
\Trace\measurex\,(\rb\star n^{\mu}\Amu -\barc\star n^{\mu}\rD_{\mu}[\A]\rc),
\eqno\numeq
$$\namelasteq\action
where $\rD_{\mu}[\A]\rc=\partial_\mu \rc- i(\Amu\star\rc-\rc\star\Amu) $.
The fields $\Amu$, $\rb$, $\barc$  and $\rc$ take values in the Lie 
algebra of $U(N)$. We define the generators of this Lie algebra as
$N\times N$ hermitian matrices, $\{(T^a)^i_{\,j}\}_{a=0,...,N^2-1},\;i,j=1,..,N$, which satisfy $[T^a,T^b]=-if^{abc}T^c$ and which are normalized so that
$\Trace\, T^a T^b = {1\over 2} \delta^{ab}$, if $a, b \ge 1$, and 
$T^0={1\over \sqrt{2N}}$. For this normalization of the generators, $T^a$,  
the path integral is invariant under arbitrary gauge transformations only if
$k$ is an integer~\cite{\integerk} --a result which also holds for the 
noncommutative torus~\cite{\Krajewski}.

The action in eq. \action\ is invariant under BRS, $s$, and antiBRS, 
$\bar s$,  transformations:
$$
\eqalign{
&s\Amu(x)=D_{\mu}c(x),\; 
s{\bar c}(x)=\rb,\; s\rb(x)=0,\; s\rc(x)=i(\rc\star \rc)(x);\cr
&\bars \Amu(x)=D_{\mu}\barc(x),\; 
\bars\rc(x)=b,\; \bars\rb(x)=0,\; s\barc(x)=i(\barc\star \barc)(x).\cr 
}
$$
As in the commutative space-time case, the action in eq. \action\ is also 
invariant under the following infinitesimal transformations
$$
\eqalignno{
&v_\mu\Anu(x)={4\pi \over k}\epsilon_{\mu\nu\rho}n^{\rho}\barc(x),\; 
v_\mu\barc (x)=0,\; v_\mu\rb(x)=\partial_\mu\barc (x) ,\; 
v_\mu\rc(x)=\Amu(x);\cr
&{\bar v}_\mu\Anu(x)={4\pi \over k}\epsilon_{\mu\nu\rho}n^{\rho}\rc(x),\; 
{\bar v}_\mu\rc (x)=0,\;{\bar  v}_\mu\rb(x)=\partial_\mu\rc (x) ,\; 
{\bar v}_\mu\barc(x)=\Amu(x).&\numeq\cr
}\namelasteq\susy
$$
The BRS transformations, the transformations generated by $v_\mu$ and the
operator $\partial_\mu$ close upon imposing the equations of motion; thus 
generating the following trivial noncommutative generalization of the 
on-shell $N=1$ supersymmetry
algebra over commutative space-time of ref. \cite{\Piguet}: 
$$
s^2\,=\,0,\;\{v_\mu,v_\nu\}\,=\,0,\;\{s,v_\mu\}\,=\,\partial_\mu .
$$
The infinitesimal transformations generated by ${\bar v}_\mu$ were introduced
in ref.~\cite{\cpmaxial}, but they do not lead to an on-shell closed algebra.

To make sense out of the Feynman diagrams which yield the 1PI functional of
the theory with action in eq.~\action, we need to find first our way around 
the axial IR singularity, $pn=0$, of the free propagators, i.e., we need to 
define  the object 
$$
{1\over pn}
$$
as a distribution. This we shall do by using the generalized 
Leibbrandt-Mandelstam prescription:
$$ 
{1\over pn}\,\equiv\,{pn^* \over pn\,pn^* + i0^+},\eqno\numeq
$$\namelasteq\LMpres
where $n^*$ is a suitable  vector. If space-time is commutative, the
generalized Leibbrandt-Mandelstam prescription allows, unlike the 
principal-value prescription,  Wick rotation of 
the Green functions whatever the axial-type gauge chosen 
--the light-cone gauge, in particular-- and whatever the  space-time 
dimension. The  prescription in eq.~\LMpres\ can thus 
be used to define Green functions whose UV finiteness may be settled by 
using power-counting arguments~\cite{\Power}. This property also holds 
for theories defined on noncommutative Minkowski space-times of electric 
type --and light-like type-- since the Moyal phases of the interaction 
vertices do not preclude Wick rotation to Euclidean noncommutative  
space-time~\cite{\NCWick}. 
In this paper we shall not need to use any particular particular 
realization of $n^*$, all we shall use are the aforementioned properties of the
Leibbrandt-Mandelstam prescription. The reader is referred to 
refs.~\cite{\LM},  \cite{\GLM} and \cite{\axialbooks} for further 
information on the generalized Leibbrandt-Mandelstam prescription.    

Now that we have defined the free propagators, we may go on and check whether
 loop corrections to the Green functions are UV finite or they give rise to UV 
divergences. The same type of power-counting arguments as in the ordinary 
case \cite{\Power, \cpmaxial} can be applied to the planar contribution of 
the Green functions to conclude that the noncommutative  theory presents 
UV divergences.
To regulate these divergences we shall use the family of regulators introduced
in ref.~\cite{\cpmaxial}. These regularizations render absolutely 
convergent the Feynman diagrams of the theory upon Wick rotation to 
Euclidean noncommutative space-time. Furthermore, they preserve explicitly 
the infinitesimal symmetries in eq.~\susy, for the regularized action reads: 
$$ 
\eqalignno{ S_\Lambda=&-{k\over 4\pi}\Trace\measurex\,\levicivita\Bigl[
\Amu\star\Reg\partirho\Anu
-{2i \over 3}\Amu\star\Arho\star\Anu\Bigr]\cr
&+\,2\,
\Trace\measurex\,\Bigl[\rb\star\Reg n^{\mu}\Amu 
-\barc\star\Reg n^{\mu}\partial_\mu\rc\,+i\barc\star\,(\Amu\star\rc-\rc\star\Amu)\Bigr].&\numeq\cr
}\namelasteq\regaction
$$
$\Reg$ is any of the functions below: 
$$
\Reg\,=\,e^{-{\partial^2\over\Lambda^2}},\quad
\Reg\,=\,\Bigl(1+{\partial^2\over\Lambda^2}\Bigr)^m,\; m\geq 2, 
m\;{\rm an}\;{\rm integer}.
$$
The regularized action in eq.~\regaction\ is not BRS invariant 
--neither  antiBRS invariant. However, we shall show that 
$$
\Gamma_\Lambda\,=\,S_\Lambda,\eqno\numeq
$$\namelasteq\noquantumcorr
$\Gamma_\Lambda$ being the 1PI functional for the action in eq.~\regaction, 
so that the limit 
$$
\lim_{\Lambda\to\infty}\,\Gamma_\Lambda\,=\, S,\eqno\numeq
$$\namelasteq\susyaction
$S$ being the action in eq.~\action, yields a renormalized theory with no 
loop corrections to the tree-level 1PI functional. Obviously, the renormalized
theory thus defined is invariant under 	BRS and antiBRS transformations, let
alone the transformations in eq.~\susy. Notice that for the regularizations
chosen, which explicitly preserve the infinitesimal vector symmetries in 
eq.~\susy, no shift of the Chern-Simons parameter $k$ occurs. The 
regularization method used in ref.~\cite{\Leibmart} will yield, presumably,
 such a shift, at the cost of spoiling the  supersymmetry invariance of the 
theory defined by eq.~\susyaction --see ref.~\cite{\Leibchen} for the 
``commutative'' analysis.  

The Feynman rules for the action in eq.~\regaction\ are given in Fig. 1, 
where  
$$
{\rm V}^{i_1i_2 i_3}_{j_1 j_2 j_3}(k_1,k_2)\,=\,
                \delta^{i_1}_{j_3}\delta^{i_2}_{j_1}\delta^{i_3}_{j_2} 
                     \;e^{{i\over 2}\theta(k_1,k_2)} -
                     \delta^{i_1}_{j_2}\delta^{i_2}_{j_3}\delta^{i_3}_{j_1} 
                     \;e^{-{i\over 2}\theta(k_1,k_2)}.
$$

\topinsert

{\settabs 4\columns \def\graphwidth{1.26in} 
\eightpoint

\+\hfil$\vcenter{\epsfxsize=\graphwidth\epsffile{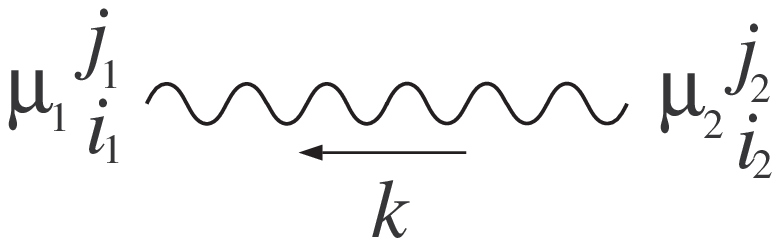}}$\hfil
 & $\vcenter{
      \hbox{$\qquad-{2\pi\over k} 
  \;\delta_{i_1}^{j_2} \delta_{i_2}^{j_1}\; \epsilon_{\mu_1\rho\mu_2}n^\rho\;
{1\over {\rm R}(p^2/\Lambda^2)}\;{1\over pn}$}
     }$\cr
\bigskip
\medskip

\+\hfil$\vcenter{\epsfxsize=\graphwidth\epsffile{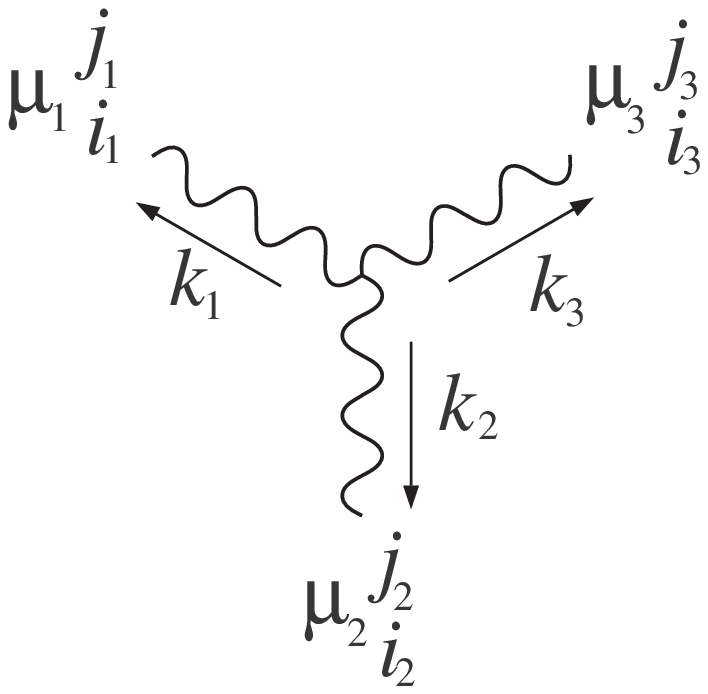}}$\hfil
 & $\vcenter{
      \hbox{$\qquad{k\over 2\pi}\;\epsilon^{\mu_1\mu_2\mu_3}\;
{\rm V}^{i_1i_2 i_3}_{j_1 j_2 j_3}(k_1,k_2)
            $}
      }$\cr
\bigskip
\medskip
\bigskip
\medskip

\+\hfil$\vcenter{\epsfxsize=\graphwidth\epsffile{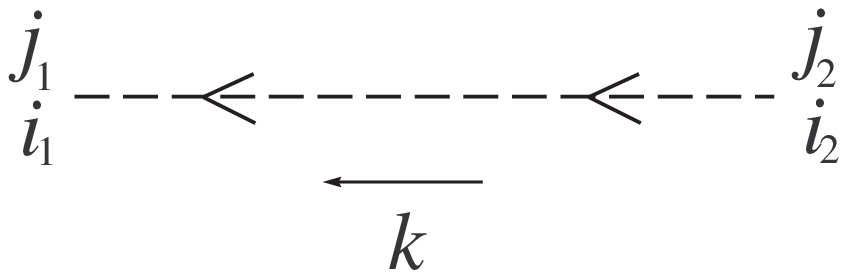}}$\hfil
  & $\vcenter{
        \hbox{$\qquad {1\over 2}
  \;\delta_{i_1}^{j_2} \delta_{i_2}^{j_1}\;
{1\over {\rm R}(p^2/\Lambda^2)}\;{1\over pn} 
              $}
   }$\cr
\bigskip
\medskip

\+\hfil$\vcenter{\epsfxsize=\graphwidth\epsffile{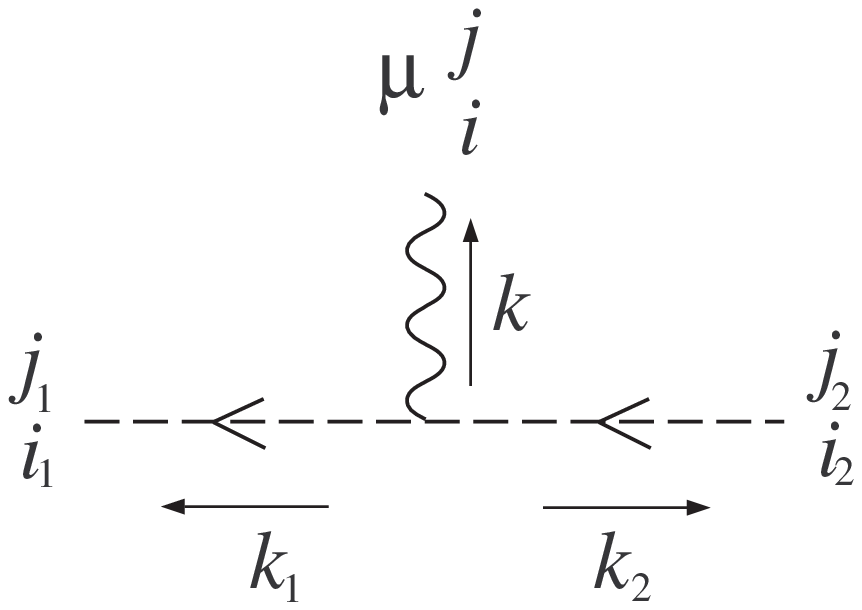}}$\hfil
  & $\vcenter{
        \hbox{$\qquad n^{\mu}\;{\rm V}^{i_1i_2 i}_{j_1 j_2 j}(k_1,k_2)
              $}
    }$\cr
}
\vskip 12pt
\narrower\noindent {\bf Figure 1.-}
{\eightpoint Feynman rules for noncommutative  $U(N)$ Chern-Simons theory in
a homogeneous axial gauge. Note that $ 
              {\rm V}^{i_1i_2 i_3}_{j_1 j_2 j_3}(k_1,k_2)\,=\,
                \delta^{i_1}_{j_3}\delta^{i_2}_{j_1}\delta^{i_3}_{j_2} 
                     \;e^{{i\over 2}\theta(k_1,k_2)} -
                     \delta^{i_1}_{j_2}\delta^{i_2}_{j_3}\delta^{i_3}_{j_1} 
                     \;e^{-{i\over 2}\theta(k_1,k_2)}.$
}

\vskip 0.1cm
\endinsert

Let us move on and show that eq.~\noquantumcorr\ holds indeed. Let us call a 
vertex of a 1PI Feynman diagram internal if none of its legs is 
an external leg.
Every 1PI Feynman diagram with two loops or more contains at least one internal
vertex. An internal vertex can be of A-A-A--type or of $\barc$-A-c--type.
The contribution to the diagram coming from each of these vertices with their 
legs on reads
$$ 
\eqalign{& -{\rm V}_{i_1i_2 i_3}^{j_1 j_2 j_3}(k_1,k_2)
\prod_{i=1}^{3}\;{1\over R(k_i^2/\Lambda^2)}\;
{k_i n^{*}\over k_i n\,k_i n^{*}+i0^+}\,\epsilon^{\nu_1\nu_2\nu_3}\,
\epsilon_{\mu_1\rho_1\nu_1}\epsilon_{\mu_2\rho_2\nu_2}
\epsilon_{\mu_3\rho_3\nu_3}\,n^{\rho_1}\,n^{\rho_2}\,n^{\rho_3},\cr 
&-{\pi \over 2k}\, {\rm V}_{i_1i_2 i}^{j_1 j_2 j}(k_1,k_2)
\prod_{i=1}^{3}\;{1\over R(k_i^2/\Lambda^2)}\;
{k_i n^{*}\over k_i n\,k_i n^{*}+i0^+}\,
\epsilon_{\mu\rho\sigma}\,n^\rho\,n^\sigma
,\cr
}
$$
respectively. The second contribution obviously vanishes and so it does the
first one since $\epsilon^{\nu_1\nu_2\nu_3}\epsilon_{\mu_2\rho_2\nu_2}
\epsilon_{\mu_3\rho_3\nu_3}\,n^{\rho_2}\,n^{\rho_3}=n^{\nu_1}
\epsilon_{\mu_2\mu_3\sigma}n^{\sigma}$. We have thus shown that beyond one-loop
there are no contributions to the 1PI functional of the regularized theory 
defined by the action in eq.~{\regaction}. Now for the one-loop loop 
contributions.

Let 
$\Gamma^{i_1\cdots i_{\rm E}\;\mu_1\cdots\mu_{\rm E}}_{j_1\cdots j_{\rm E}}
(k_1,\cdots, k_{\rm E})$ denote the one-loop contribution to the 1PI Green 
function of ${\rm E}$ gauge fields and no ghost field.  If ${\rm E}\geq 3$,
then 
$$
\Gamma^{i_1\cdots i_{\rm E}\;\mu_1\cdots\mu_{\rm E}}_{j_1\cdots j_{\rm E}}
(k_1,\cdots, k_{\rm E})\,=\,\Gamma^{({\rm gauge}\;{\rm modes})}\,+\,
\Gamma^{({\rm clockwise} )} \,+\,\Gamma^{({\rm anticlockwise} )},\eqno\numeq
$$\namelasteq\higherlegs
where $\Gamma^{({\rm gauge}\;{\rm modes})}$, $\Gamma^{({\rm clockwise} )}$ 
and $\Gamma^{({\rm anticlockwise} )}$ are, respectively, the contributions
furnished by gauge modes going around the loop and ghost modes propagating 
around the loop clockwise and anticlockwise. Using the Feynman rules in 
fig. 1, one can readily show that
$$ 
\eqalignno{&\Gamma^{({\rm gauge}\;{\rm modes})}\,=\,
\measureq\;\prod_{i=1}^{\E}{1\over R(p_i^2/\Lambda^2)}\;
{p_i n^{*}\over p_i n\,p_i n^{*}+i0^+}\prod_{i=1}^{\E}
 {\rm V}^{i_i m_{i-1} l_i}_{j_i l_{i-1} m_i}(-k_i,p_{i-1})
\prod_{i=1}^{\E}{\rm N}^{\mu_i\nu_{i-1}}_{\nu_i},\cr
&\Gamma^{({\rm clockwise} )}\,=\,-(-1)^{{\E}}\prod_{i=1}^{\E} n^{\mu_i}\,
\measureq\;\prod_{i=1}^{\E}{1\over R(p_i^2/\Lambda^2)}\;
{p_i n^{*}\over p_i n\,p_i n^{*}+i0^+}\prod_{i=1}^{\E}
 {\rm V}^{i_i m_{i-1} l_i}_{j_i l_{i-1} m_i}(-p_{i-1},p_i)\cr
&{\rm and}\cr
&\Gamma^{({\rm anticlockwise} )}\,=\,-(-1)^{{\E}}\prod_{i=1}^{\E} n^{\mu_i}\,
\measureq\;\prod_{i=1}^{\E}{1\over R(p_i^2/\Lambda^2)}\;
{p_i n^{*}\over p_i n\,p_i n^{*}+i0^+}\prod_{i=1}^{\E}
 {\rm V}^{m_{i-1} l_i i_i}_{l_{i-1} m_i j_i}(-p_{i-1},p_i).\cr
& &\numeq\cr
}\namelasteq\results
$$
In the previous equation $l_0$, $m_0$, $v_0$ and $p_0$ are equal to 
$l_\E$, $m_\E$, $v_\E$ and $p_\E$, respectively, and $p_1=q$ and 
$k_i-p_{i-1}+p_i=0$. $k_i$, $-p_{i-1}$ and $p_i$ are the momenta coming out
of the vertex with index $i$. Further, the symbol 
${\rm N}^{\mu_i\nu_{i-1}}_{\nu_i}$ denotes the tensor contraction 
$-\epsilon^{\mu_i\nu_{i-1}\rho_i}\epsilon_{\rho_i\sigma\nu_i}n^{\sigma}$.

Now, it is not difficult to show that 
$$
\prod_{i=1}^{\E}{\rm N}^{\mu_i\nu_{i-1}}_{\nu_i}\,=\,2\,
(-1)^{\E}\,\prod_{i=1}^{\E} n^{\mu_i},\quad
{\rm V}^{l_i m_{i-1} i_i}_{m_i l_{i-1} j_i}(p_i, -p_{i-1})\,=\,
-\,{\rm V}^{i_i m_{i-1} l_i}_{j_i l_{i-1} m_i}(-p_{i-1},p_i)
$$
and
$$
{\rm V}^{i_i m_{i-1} l_i}_{j_i l_{i-1} m_i}(-p_{i-1},p_i)\,=\,
{\rm V}^{i_i m_{i-1} l_i}_{j_i l_{i-1} m_i}(-k_i,p_{i-1}),\;
{\rm V}^{m_{i-1} l_i i_i}_{l_{i-1} m_i j_i}(-p_{i-1},p_i)\,=\,
{\rm V}^{i_i m_{i-1} l_i}_{j_i l_{i-1} m_i}(-k_i,p_{i-1});
$$
which employed along with  eq.~\results\ yield
$$
\Gamma^{({\rm clockwise} )} \,=\,\Gamma^{({\rm anticlockwise} )}\,=\,
-{1\over 2}\,\Gamma^{({\rm gauge}\;{\rm modes})}.
$$
If we substitute  this result in eq.~\higherlegs, we will conclude that 
$$
\Gamma^{i_1\cdots i_{\rm E}\;\mu_1\cdots\mu_{\rm E}}_{j_1\cdots j_{\rm E}}
(k_1,\cdots, k_{\rm E})\,=\, 0,
$$
whenever $\E\geq 2$. This result also holds for $\E\,=\, 1$, $2$ since in both
these cases
$$
\Gamma^{i_1\cdots i_{\rm E}\;\mu_1\cdots\mu_{\rm E}}_{j_1\cdots j_{\rm E}}
(k_1,\cdots, k_{\rm E})\,=\,{1\over 2}\,\Gamma^{({\rm gauge}\;{\rm modes})}\,+\,
\Gamma^{({\rm clockwise} )}, 
$$\namelasteq\twolegs
with $\Gamma^{({\rm gauge}\;{\rm modes})}$ and $\Gamma^{({\rm clockwise} )}$ 
given in eq.~\results. Notice that the factor ${1\over 2}$ on the r.h.s of 
eq.~\twolegs\ is the symmetry factor of the diagram with gauge modes  
propagating around the loop. 
 
The one-loop contribution to any 1PI Feynman diagram with ghost vanishes since
it has at least an internal gauge propagator  contracted with the 
$n^\mu$ vector of a  ${\bar c}$-A-c vertex, a contraction which vanishes.

\bigskip

\section{3.- Conclusions and comments}

We have shown in this paper that $U(N)$ Chern-Simons theory on the 
noncommutative plane can be defined in perturbation theory so that 
its 1PI functional in an arbitrary homogeneous axial gauge receives 
no quantum corrections. To construct the theory we have used, as an 
intermediate stage, a regularization method which preserves the infinitesimal 
vector Grassmann symmetry of the theory. 

Let us  state next that for our definition of the theory the following 
results hold
$$
\eqalignno{&\langle\prod_{l=1}^{2\E+1}\A^{i_l}_{j_l\,\mu_l}(x_l)\rangle\,=\,0,
\cr	
&\langle\prod_{l=1}^{2\E}\A^{i_l}_{j_l\,\mu_l}(x_l)\rangle\,=\,
\sum_{{\rm P}}\,\prod_{m=1}^{\E}\langle \A^{i_{l_m}}_{j_{l_m}\,\mu_{l_m}}(x_{l_m})
\A^{i_{n_m}}_{j_{n_m}\,\mu_{n_m}}(x_{n_m})\rangle,&\numeq\cr
}
$$\namelasteq\pairings
where ${\rm P}\,=\,\{(l_1,n_1),(l_2,n_2),\cdots,(l_{\E},n_{\E})\}$ is the
set of all possible collections of $\E$ pairings of the elements of
$\{1,\cdots\,\E\}$ with ${l_m} < {n_m}$. To prove the results in eq.~\pairings,
one may use the invariance of the theory under the first infinitesimal 
transformation in eq.~\susy, the fact that by ghost number conservation
$$
\langle\prod_{l=1}^{\E}\A^{i_l}_{j_l\,\mu_l}(x_l)
\rc^{i_{\E+1}}_{j_{\E+1}}(x_{\E+1})\rangle\,=\,0,
$$
and the fact that for our definition of the noncommutative theory 
$$
\langle {\rm O}_1(\A){\rm O}_2(\rc\barc)\rangle\,=\,
\langle {\rm O}_1(\A)\rangle\langle
{\rm O}_2(\rc\barc)\rangle.
$$
Hence, the Green functions of the elementary fields of the noncommutative 
theory are in perturbation theory those of the ordinary (over commutative 
space-time) theory. Of course, as happens in the ordinary case, it is the 
vev of gauge invariant operators which carries the really nontrivial 
information --barring the famous one-loop shift of $k$. We will report 
elsewhere on the equations satisfied by  vev of Wilson lines and the 
connection of noncommutative Chern-Simons theory with the noncommutative 
Wess-Zumino-Witten model~\cite{\WZW}.

Finally, when I was writing this paper, I  came across the paper quoted in 
ref.~\cite{\DasJab} where radiative corrections in Noncommutative Chern-Simons 
theory in an axial space-like gauge are analyzed. In the latter paper 
the principal value prescription is used and the results therein displayed
agree with ours for space-like gauges.

\section{Acknowledgments} 
This work has been partially supported by CICyT under grant PB98-0842.
As I was writing the closing sentences of this paper I was informed
of the demise of Professor George Leibbrandt; to him I dedicate this paper
with all my everlasting gratitude. Requiescat In Pace.

\section{References}

\frenchspacing

\refno\TMYM.
S. Deser, R. Jackiw and S. Templeton, Ann. Phys. {\bf 140} (1982) 372.

\refno\Wittenjones.
E. Witten, Comm. Math. Phys. {\bf 121} (1989) 351.

\refno\HALL.
R. Mackenzie and F. Wilczek, Int. J. Mod. Phys. {\bf A3} (1988) 2827;
Y.H. Chen, F. Wilczek, E. Witten and B. I. Halperin, Int. J. Phys. {\bf B3}
(1989) 1001; A. Lopez and E. Fradkin, ``{\it Fermionic Chern-Simons Theory for
the Fractional Hall Effect}'' in Composite Fermions in the Quantum Hall Effect,
edited by O, Heinonen, {\tt cond-Mat/9704055}; S.H. Simon, 
``{\it The Chern-Simons Fermi Liquid Description of the 
Quantum Hall States}'' in Composite Fermions", 
ed. O. Heinonen, World Scientific, {\tt cond-Mat/9812186};
G.V. Dunne, ``{\it Aspects of Chern-Simons Theory}'', lectures in 
Topological Aspects of Low Dimensional Systems, Les Houches Summer School 1998,
{\tt hep-th/9902115}.

\refno\Labastida.
J.M.F. Labastida, ``{\it Chern-Simons Gauge Theory: Ten Years After}'',
lecture delivered at the workshop ``Trends in Theoretical Physics II'', Buenos Aires, November 1998, {\tt hep-th/9905057}.

\refno\Vassinv.
J.M.F. Labastida and  Esther Perez, J.Math.Phys. {\bf 39} (1998) 
5183-5198;  J.Math.Phys. {\bf 41} (2000) 2658-2699.

\refno\King.
S. King and J. Fr\"olich, Comm. Math. Phys. {\bf 126} (1989) 167.

\refno\PP.
J.M. Gracia-Bond\'{\i}a, J.C. V\'arilly and H. Figueroa, ``{\it Elements of 
Noncommutative Geometry}'', Birk\"auser, 2001. 

\refno\Krajewski.
A. Chamseddine and J. Fr\"olich, J. Math. Phys. {\bf 35} (1994) 5195;
T. Krajewski, ``{\it Gauge invariance of the Chern-Simons action in noncommutative geometry}'', {\tt  math-ph/990}\par {\tt 3047}.

\refno\NCCSper.
A. A. Bichl, J. M. Grimstrup, V. Putz and  M. Schweda, JHEP {\bf 0007} 
(2000) 046; Guang-Hong Chen and Yong-Shi Wu, Nucl.Phys. {\bf B593} (2001) 562.

\refno\NCCSnoper.
A. P. Polychronakos, JHEP {\bf 0011} (2000) 008;
J. Kluson, ``{\it Matrix model and Chern-Simons theory}'', 
{\tt hep-th/0012184}; M.M. Sheikh-Jabbari, 
``{\it A Note on Noncommutative Chern-Simons Theories}'',
 {\tt hep-th/0102092};
Dongsu Bak, Sung Ku Kim, Kwang-Sup Soh and Jae Hyung Yee, 
``{\it Noncommutative Chern-Simons Solitons}'', {\tt hep-th/0102137};

\refno\integerk.
V.P. Nair and A.P. Polychronakos, 
``{\it On Level Quantization for the Noncommutative Chern-Simons Theory}'', 
{\tt hep-th/0102181};
D. Bak, K. Lee and J.-H. Park, ``{\it Chern-Simons Theories on the 
Noncommutative Plane}'', {\tt hep-th/0102188}.

\refno\induced.
C.-S. Chu, Nucl. Phys. {\bf B580} (2000) 352; N. Grandi and  G.A. Silva, 
``{\it   Chern-Simons action in noncommutative space}'', {\tt hep-th/0010113}.

\refno\NCHALL.
L. Susskind, ``{\it The quantum Hall fluid and non-commutative Chern-Simons 
theory}'', {\tt hep-th/0101029}; A.P. Polychronakos, ``{\it Quantum Hall 
states as matrix Chern-Simons theory}'', {\tt hep-th/0103013};   
S. Hellerman and M. Van Raamsdonk, ``{\it Quantum Hall Physics Equals  
Noncommutative Field Theory}'', {\tt  hep-th/0103179}.

\refno\Piguet.
A. Brandhuber, M. Langer, M. Schweda, O. Piguet and S.P. Sorella, 
Phys. Lett. {\bf B300} (1993) 92.

\refno\cpmaxial.
C.P. Martin, Phys. Lett. {\bf B263} (1991) 69.

\refno\Power.
A. Basetto, M. Dalbosco and R. Soldati, Phys. Rev. {\bf D36} (1987)3138.

\refno\NCWick.
Jaume Gomis and T.  Mehen, Nucl.Phys. B591 (2000) 265-276;
O. Aharony, Jaume Gomis and T. Mehen, JHEP {\bf 0009} (2000) 023;
L. Alvarez-Gaum\'e, J.L.F. Barb\'on and  R. Zwicky, 
``{\it Remarks on Time-Space Noncommutative Field Theories}'', 
{\tt hep-th/0103069}.

\refno\LM.
S. Mandelstam, Nucl. Phys. {\bf B213} (1983) 149; G. Leibbrandt, Phys. Rev. 
{\bf D29} (1984) 1699.

\refno\GLM.
A. Bassetto, M. Dalbosco, I. Lazzizera  and R. Soldati,  Phys. Rev. 
{\bf D31} (1985) 2012;
P. Gaigg, M. Kreuzer, M. Schweda and O. Piguet, J. Math. Phys. {\bf 28} (1987)
2781; P. Gaigg and M. Kreuzer, Phys. Lett. {\bf B205} (1988) 530;
G. Leibbrandt, Nucl. Phys. {\bf B320} (1988) 405; 
I. Lazzizera, Phys. Lett. {\bf B210} (1988) 188;
H. H\"uffel, P.V. Landshoff 
and J.C. Taylor, Phys. Lett. {\bf B217} (1989) 147; P.V. Landshoff,
Phys. Lett. {\bf B227} (1989) 427.

\refno\axialbooks.
A. Basetto, G. Nardelli and R. Soldati, ``{\it Yang-Mills theories  
in algebraicnon-covariant gauges}'', World Scientific 1991; G. Leibbrandt,   
``{\it Noncovariant Gauges}'', World Scientific, 1994.

\refno\Leibmart.
G. Leibbrandt and C.P. Martin, Nucl. Phys. {\bf B377} (1992) 593.

\refno\Leibchen.
W.F. Chen and G. Leibbrandt, ``{\it Vector Supersymmetry and Finite Quantum 
Correction of Chern-Simons Theory in the Light-Cone Gauge}'', {\tt hep-th/0012173}.

\refno\WZW.
E.F. Moreno and F.A. Schaposnik, JHEP {\bf 0003} (2000) 032; 
K. Furuta and T. Inami, Mod. Phys. Lett. {\bf A15} (2000) 997; 
E.F. Moreno and F.A. Schaposnik, Nucl. Phys. {\bf B596} (2001) 439-458;
A.M. Ghezelbash and S. Parvizi, Nucl. Phys. {\bf B592} (2001) 408-416;
A.R. Lugo, ``{\it Correlation functions in the non-commutative Wess-Zumino-Witten model}'', {\tt hep-th/0012268 }.

\refno\DasJab.
A. Das and M.M. Sheikh-Jabbari, 
{\it ``Absence of Higher Order Corrections to Noncommutative Chern-Simons
Theory Coupling}'', {\tt hep-th/0103139}.

\bye